\documentclass[
	conference,
	10pt
]{IEEEtran}

\IEEEoverridecommandlockouts 
\makeatletter
\def\markboth#1#2{\def\leftmark{\@IEEEcompsoconly{\sffamily}\MakeUppercase{\protect#1}}%
\def\rightmark{\@IEEEcompsoconly{\sffamily}\MakeUppercase{\protect#2}}}
\makeatother

\usepackage[amssymb]{SIunits}
\usepackage[cmex10]{amsmath} 
\usepackage{amssymb, amsthm}
\usepackage{bbm} 
\usepackage[latin1]{inputenc}
\usepackage{xcolor}
\usepackage{cite}
\usepackage[pdftex]{graphicx}
\usepackage[caption=false, font=footnotesize]{subfig}
\usepackage{url}
\usepackage{mathtools}

\usepackage{xspace} 

\usepackage{booktabs} 

\usepackage{array}
\newcolumntype{L}[1]{>{\raggedright\let\newline\\\arraybackslash\hspace{0pt}}m{#1}}
\newcolumntype{C}[1]{>{\centering\let\newline\\\arraybackslash\hspace{0pt}}m{#1}}
\newcolumntype{R}[1]{>{\raggedleft\let\newline\\\arraybackslash\hspace{0pt}}m{#1}}

\usepackage[
	shortcuts,
	nonumberlist,	
	acronym, 
	hyperfirst=true,
	section=section, 
	order=letter,
	numberedsection=nolabel 
]{glossaries}

\theoremstyle{definition}

\theoremstyle{plain}

\theoremstyle{remark} 

\newtheorem{example}{Example}

\newcommand\xqed[1]{%
\leavevmode\unskip\penalty9999 \hbox{}\nobreak\hfill
\quad\hbox{#1}}
\newcommand\demo{\xqed{$\triangle$}}

\newcommand{\define}{\triangleq}

\newcommand{\E}[1]{\mathbb{E}\left[#1\right]}

\newcommand{\Aunif}{\mat{A}_\text{uni}}

\newcommand{\tee}{\ensuremath{t}}

\newcommand{\Gpc}{\ensuremath{\mathcal{C}_n(\etab, \vect{\tau})}}

\newcommand{\tmax}{\ensuremath{\tee_{\text{max}}}}

\newcommand{\transpose}{\intercal}

\newcommand{\cthr}{\ensuremath{\bar{c}}}
\newcommand{\vect}[1]{\ensuremath{\boldsymbol{#1}}}
\newcommand{\mat}[1]{\ensuremath{\boldsymbol{#1}}}
\newcommand{\etab}{\ensuremath{\boldsymbol{\eta}}}

\newcommand{\ch}{M} 
\newcommand{\snr}{\rho} 

\newif\ifshow
\showfalse

\newcommand{\abbr}[1]{{#1}}				

\makeatletter
\let\aclOLD=\acl
\renewcommand{\acl}[1]{%
  \begingroup    
  \let\@@underline=\relax
  \aclOLD{#1}%
  \endgroup
}
\makeatother

\newcommand{\NewA}[3]{
	\newacronym{#1}{#2}{#3}
}

\NewA{af}{AF}{\abbr{a}mplify-and-\abbr{f}orward}
\NewA{apsk}{APSK}{\abbr{a}mplitude \abbr{p}hase-\abbr{s}hift \abbr{k}eying}
\NewA{ask}{ASK}{\abbr{a}mplitude-\abbr{s}hift \abbr{k}eying}
\NewA{ase}{ASE}{\abbr{a}mplified \abbr{s}pontaneous \abbr{e}mission}
\NewA{awgn}{AWGN}{\abbr{a}dditive \abbr{w}hite \abbr{G}aussian \abbr{n}oise}
\NewA{biawgn}{BI-AWGN}{binary-input \abbr{a}dditive \abbr{w}hite \abbr{G}aussian \abbr{n}oise}
\NewA{bep}{BEP}{\abbr{b}it \abbr{e}rror \abbr{p}robability}
\NewA{cep}{CEP}{\abbr{c}odeword \abbr{e}rror \abbr{p}robability}
\NewA{ber}{BER}{\abbr{b}it \abbr{e}rror \abbr{r}ate}
\NewA{qap}{QAP}{\abbr{q}uadratic \abbr{a}assignment \abbr{p}roblem}
\NewA{bicm}{BICM}{\abbr{b}it-\abbr{i}nterleaved \abbr{c}oded \abbr{m}odulation}				
\NewA{cm}{CM}{\abbr{c}oded \abbr{m}odulation}				
\NewA{qpsk}{QPSK}{quadrature phase-shift keying}				
\NewA{mlcm}{MLCM}{multilevel coded modulation}				
\NewA{bpsk}{BPSK}{\abbr{b}inary \abbr{p}hase-\abbr{s}hift \abbr{k}eying}				
\NewA{bsc}{BSC}{\abbr{b}inary \abbr{s}ymmetric \abbr{c}hannel}				
\NewA{brgc}{BRGC}{\abbr{b}inary \abbr{r}eflected \abbr{G}ray \abbr{c}ode}				
\NewA{cf}{CF}{\abbr{c}haracteristic \abbr{f}unction}
\NewA{csit}{CSIT}{\abbr{c}annnel \abbr{s}tate \abbr{i}nformation at the \abbr{transmitter}}
\NewA{csi}{CSI}{\abbr{c}annnel \abbr{s}tate \abbr{i}nformation}
\NewA{df}{DF}{\abbr{d}ecode-and-\abbr{f}orward}
\NewA{fd}{FD}{\abbr{f}ull-\abbr{d}uplex}
\NewA{fft}{FFT}{\abbr{f}ast \abbr{F}ourier \abbr{t}ransform}
\NewA{iid}{IID}{independent and \abbr{i}dentically \abbr{d}istributed}
\NewA{isi}{ISI}{\abbr{i}nter\abbr{s}ymbol \abbr{i}nterference}
\NewA{lb}{LB}{\abbr{l}ower \abbr{b}ound}
\NewA{map}{MAP}{\abbr{m}aximum \abbr{a} \abbr{p}osteriori}
\NewA{mf}{MF}{\abbr{m}odulo-and-\abbr{f}orward}
\NewA{mlan}{MLAN}{\abbr{m}odulo-\abbr{l}attice \abbr{a}dditive \abbr{n}oise}
\NewA{ml}{ML}{\abbr{m}aximum \abbr{l}ikelihood}
\NewA{mmse}{MMSE}{\abbr{m}inimum \abbr{m}ean \abbr{s}quare \abbr{e}rror}
\NewA{nlpc}{NLPC}{\abbr{n}onlinear \abbr{p}hase \abbr{c}ompensation}
\NewA{nlpn}{NLPN}{\abbr{n}onlinear \abbr{p}hase \abbr{n}oise}
\NewA{nc}{NC}{\abbr{n}etwork \abbr{c}oding}
\NewA{ofmd}{OFDM}{\abbr{o}rthogonal \abbr{f}requency-\abbr{d}ivision \abbr{m}ultiplexing}			
\NewA{dp}{DP}{\abbr{d}ual-\abbr{p}olarization}
\NewA{pam}{PAM}{\abbr{p}ulse \abbr{a}mplitude \abbr{m}odulation}
\NewA{pdf}{PDF}{\abbr{p}robability \abbr{d}ensity \abbr{f}unction}
\NewA{plnc}{PNC}{\abbr{p}hysical-layer \abbr{n}etwork \abbr{c}oding}
\NewA{psk}{PSK}{\abbr{p}hase-\abbr{s}hift \abbr{k}eying}
\NewA{pmd}{PMD}{\abbr{p}olarization \abbr{m}ode \abbr{d}ispersion}
\NewA{pm}{PM}{\abbr{p}olarization-\abbr{m}ultiplexed}
\NewA{pdm}{PDM}{\abbr{p}olarization-\abbr{d}ivision \abbr{m}ultiplexing}
\NewA{qam}{QAM}{\abbr{q}uadrature \abbr{a}mplitude \abbr{m}odulation}
\NewA{sqp}{SQP}{\abbr{s}equential \abbr{q}uadratic \abbr{p}rogramming}
\NewA{rd}{RD}{\abbr{r}adii \abbr{d}istribution}
\NewA{sep}{SEP}{\abbr{s}ymbol \abbr{e}rror \abbr{p}robability}
\NewA{ser}{SER}{\abbr{s}ymbol \abbr{e}rror \abbr{r}ate}
\NewA{si}{SI}{\abbr{s}ide \abbr{i}nformation}
\NewA{sp}{SP}{\abbr{s}ingle-\abbr{p}olarization}
\NewA{sanr}{SNR}{\abbr{s}ignal-to-(additive-)\abbr{n}oise \abbr{r}atio}
\NewA{snr}{SNR}{\abbr{s}ignal-to-\abbr{n}oise \abbr{r}atio}
\NewA{snlse}{sNLSE}{\abbr{s}tochastic \abbr{n}onlinear \abbr{S}chr\"odinger \abbr{e}quation}
\NewA{nlse}{NLSE}{\abbr{n}onlinear \abbr{S}chr\"odinger \abbr{e}quation}
\NewA{stwrc}{sTRC}{\abbr{s}eparated \abbr{t}wo-way \abbr{r}elay \abbr{c}hannel}
\NewA{stwtrc}{sTTRC}{\abbr{s}eparated \abbr{t}wo-way \abbr{t}wo-\abbr{r}elay \abbr{c}hannel}
\NewA{ts}{TS}{\abbr{t}wo-\abbr{s}tage}
\NewA{twrc}{TRC}{\abbr{t}wo-way \abbr{r}elay \abbr{c}hannel}
\NewA{twtrc}{TTRC}{\abbr{t}wo-way \abbr{t}wo-\abbr{r}elay \abbr{c}hannel}
\NewA{wdm}{WDM}{\abbr{w}avelength-\abbr{d}ivision \abbr{m}ultiplexing}
\NewA{vn}{VN}{variable node}
\NewA{cn}{CN}{constraint node}
\NewA{de}{DE}{density evolution}
\NewA{sc}{SC}{spatially-coupled}
\NewA{ldpc}{LDPC}{low-density parity-check}
\NewA{bp}{BP}{belief propagation}
\NewA{dm}{DM}{dispersion-managed}
\NewA{spm}{SPM}{self-phase modulation}
\NewA{xpm}{XPM}{cross-phase modulation}
\NewA{fwm}{FWM}{four-wave-mixing}
\NewA{ixpm}{IXPM}{intrachannel cross-phase modulation}
\NewA{ifwm}{IFWM}{intrachannel four-wave-mixing}
\NewA{ssfm}{SSFM}{split-step Fourier method}
\NewA{fec}{FEC}{forward error correction}
\NewA{psd}{PSD}{power spectral density}
\NewA{ook}{OOK}{on-off keying}
\NewA{smf}{SMF}{single-mode fiber}
\NewA{ssmf}{SSMF}{standard single-mode fiber}
\NewA{dbp}{DBP}{digital backpropagation}
\NewA{edfa}{EDFA}{erbium-doped fiber amplifier}
\NewA{ofdm}{OFDM}{orthogonal frequency division multiplexing}
\NewA{exit}{EXIT}{extrinsic information transfer}
\NewA{pexit}{P-EXIT}{protograph extrinsic information transfer}
\NewA{osnr}{OSNR}{optical signal-to-noise ratio}
\NewA{roadm}{ROADM}{reconfigurable optical add-drop multiplexer}
\NewA{rps}{RPS}{Raman pump station}
\NewA{mlse}{MLSE}{maximum likelihood sequence estimation}
\NewA{dcf}{DCF}{dispersion compensating fiber}
\NewA{tcm}{TCM}{trellis coded modulation}
\NewA{edc}{EDC}{electronic dispersion compensation}
\NewA{ldpcc}{LDPCC}{low-density parity-check convolutional}
\NewA{llr}{LLR}{log-likelihood ratio}
\NewA{ra}{RA}{repeat-accumulate}
\NewA{ira}{IRA}{irregular-repeat-accumulate}
\NewA{ara}{ARA}{accumulate-repeat-accumulate}
\NewA{mi}{MI}{mutual information}
\NewA{vdmm}{VDMM}{variable degree matched mapping}
\NewA{gmi}{GMI}{generalized mutual information}
\NewA{wd}{WD}{windowed decoder}
\NewA{gn}{GN}{Gaussian noise}
\NewA{hd}{HD}{hard-decision}
\NewA{hdd}{HDD}{hard-decision decoding}
\NewA{sd}{SD}{soft-decision}
\NewA{sdd}{SDD}{soft-decision decoding}
\NewA{emp}{EMP}{extrinsic message passing}
\NewA{imp}{IMP}{intrinsic message passing}
\NewA{gldpc}{GLDPC}{generalized low-density parity-check}
\NewA{scgldpc}{SC-GLDPC}{spatially-coupled generalized low-density parity-check}
\NewA{scldpc}{SC-LDPC}{spatially-coupled low-density parity-check}
\NewA{scfec}{SC-FEC}{spatially-coupled forward error correction}
\NewA{tbbc}{TBBC}{tightly-braided block code}
\NewA{gpc}{GPC}{generalized product code}
\NewA{otn}{OTN}{optical transport network}
\NewA{bch}{BCH}{Bose--Chaudhuri--Hocquenghem}
\NewA{rs}{RS}{Reed--Solomon}
\NewA{bbbc}{BBBC}{block-wise braided block code}
\NewA{hpc}{HPC}{half-product code}
\NewA{pc}{PC}{product code}
\NewA{rv}{RV}{random variable}
\NewA{mbp}{MBP}{multiple block permutator}
\NewA{cer}{CER}{codeword error probability}
\NewA{oh}{OH}{overhead}
\NewA{bdd}{BDD}{bounded-distance decoding}
\NewA{ncg}{NCG}{net coding gain}
\NewA{gwbp}{Galton--Watson branching process}{Galton--Watson branching process}

\newacronym[%
	longplural={binary erasure channels},%
	shortplural={BECs}%
]{bec}{BEC}{binary erasure channel}%

\newcommand{\VN}{\gls{vn}\xspace}
\newcommand{\VNs}{\glspl{vn}\xspace}
\newcommand{\CN}{\gls{cn}\xspace}
\newcommand{\CNs}{\glspl{cn}\xspace}

\newcommand{\PCs}{\glspl{pc}\xspace}
\newcommand{\GPCs}{\glspl{gpc}\xspace}
\newcommand{\gPCs}{\Glspl{gpc}\xspace}
\newcommand{\HPCs}{\glspl{hpc}\xspace}
\newcommand{\HPC}{\gls{hpc}\xspace}
\newcommand{\GPC}{\gls{gpc}\xspace}
\newcommand{\PC}{\gls{pc}\xspace}

\newcommand{\BDD}{\gls{bdd}\xspace}
\newcommand{\BSC}{\gls{bsc}\xspace}
\newcommand{\BSCs}{\glspl{bsc}\xspace}
\newcommand{\BEC}{\gls{bec}\xspace}
\newcommand{\BECs}{\glspl{bec}\xspace}
\newcommand{\DE}{\gls{de}\xspace}
\newcommand{\BCH}{\gls{bch}\xspace}

\newcommand{\BICM}{\gls{bicm}\xspace}

\newcommand{\AWGN}{\gls{awgn}\xspace}

\newcommand{\SNR}{\gls{snr}\xspace}
\newcommand{\BER}{\gls{ber}\xspace}

\begin{document}

\title{Density Evolution for Deterministic Generalized Product Codes
with Higher-Order Modulation}


\author{
	\IEEEauthorblockN{
	Christian Häger\IEEEauthorrefmark{2},
	Alexandre Graell i Amat\IEEEauthorrefmark{2}, 
	Henry D.~Pfister\IEEEauthorrefmark{3}, and
	Fredrik Brännström\IEEEauthorrefmark{2} 
	\thanks{This work was partially funded by
	the Swedish Research Coucil under grant \#2011-5961. The work of H.~Pfister was
	supported in part by the National Science Foundation (NSF) under
	Grant No.~1320924. Any opinions, findings, conclusions, and
	recommendations expressed in this material are those of the authors
	and do not necessarily reflect the views of the NSF.   }}
	\IEEEauthorblockA{\IEEEauthorrefmark{2}%
	Department of Signals and Systems,
	Chalmers University of Technology,
	Gothenburg, Sweden}
	\IEEEauthorblockA{\IEEEauthorrefmark{3}%
	Department of Electrical and Computer Engineering, Duke University,
	Durham, North Carolina
	}
	}

\maketitle


\begin{abstract}
	\gPCs are extensions of \PCs where coded bits are protected by two
	component codes but not necessarily arranged in a rectangular array.
	It has recently been shown that there exists a large class of
	deterministic \GPCs (including, e.g., irregular \PCs, half-product
	codes, staircase codes, and certain braided codes) for which the
	asymptotic performance under iterative bounded-distance decoding
	over the \BEC can be rigorously characterized in terms of a density
	evolution analysis. In this paper, the analysis is extended to the
	case where transmission takes place over parallel \BECs with
	different erasure probabilities. We use this model to predict the
	code performance in a coded modulation setup with higher-order
	signal constellations. We also discuss the design of the bit mapper
	that determines the allocation of the coded bits to the modulation
	bits of the signal constellation. 
\end{abstract}
\glsresetall

\section{Introduction}


A \PC is defined as the set of all rectangular arrays such that each
row and column is a codeword in some linear component code
\cite{Elias1954}. Assuming efficient iterative decoding of the
component codes (e.g., algebraic \BDD of \BCH codes), \PCs are an
excellent choice for error-correcting codes in high-speed
applications such as fiber-optical communications \cite{Justesen2010}.
Indeed, \PCs are standardized in \cite{ITU-T2004} and several
extensions of \PCs, e.g., staircase \cite{Smith2012a} and braided
codes \cite{Jian2014}, have been proposed for such systems. We refer
to these codes as \GPCs. 

Motivated by the recent trend towards spectrally-efficient
fiber-optical systems \cite{Beygi2014}, our objective in this paper is
to characterize the asymptotic performance of \GPCs under iterative
\BDD in a coded modulation scenario. Similar to \cite{Smith2012}, we
consider pragmatic \BICM with a hard-decision symbol detector. This
setup can be modeled as a set of parallel \glspl{bsc}. The
hard-decision detector comes at the price of a performance loss
compared to calculating ``soft'' reliability information about the
coded bits.  However, it is also significantly less complex and
therefore an attractive candidate for high-speed systems.


In \cite{Haeger2015tit}, the authors propose a deterministic (i.e.,
non-ensemble-based) construction for \GPCs and study the asymptotic
performance over the \BEC under iterative \BDD in the form of a \DE
analysis. In this paper, the main contribution is to extend this
analysis to the case where transmission takes place over parallel
\BECs with different erasure probabilities. Ignoring miscorrections in
the \BDD, the analysis applies without change also to parallel \BSCs.
The derived \DE analysis can then be used to predict the waterfall
performance of the \GPCs for the considered \BICM system. 


As an application, we consider the problem of optimizing the bit
mapper (or interleaver) that determines the allocation of the coded
bits from the \GPC to the modulation bits of the signal constellation.
This problem has been studied in detail for low-density parity-check
codes (see, e.g., \cite{Haeger2014oe} and references therein for an
overview in the context of fiber-optical communications). Here, we
show that by taking advantage of the unequal error protection offered
by a higher-order signal constellation, the performance of
deterministic \GPCs can be improved at almost no increased system
complexity.


\emph{Notation.} We use boldface letters for vectors and matrices
(e.g., $\vect{x}$ and $\mat{A}$), and denote the transpose by
$(\cdot)^\transpose$. The symbols $\vect{0}$ and $\vect{1}$ denote the
all-zero and all-one vectors, where the length is apparent from the
context. For vectors, we use $\vect{x} \succeq \vect{y}$ if $x_i \geq
y_i$ for all $i$. We also define $[m] \define \{1, 2, \dots, m\}$. 


\section{Deterministic Generalized Product Codes}

In this section, we review the parametrized family of \GPCs proposed
in \cite{Haeger2015tit}. A \GPC in this family is denoted by $\Gpc$,
where $n$ corresponds to the total number of \CNs in the underlying
Tanner graph and $\etab$ is a binary symmetric $L \times L$ matrix
that defines the graph connectivity. The parameter $\vect{\tau}$ is
used to specify \GPCs that employ component codes with different
erasure-correcting capabilities and is described in Section
\ref{sec:vn_classes} below. 

To construct the Tanner graph that defines $\Gpc$, assume that there
are $L$ positions. Then, 

\begin{enumerate}
	\item place $d \define n/L$ \CNs at each position and 
		
	\item connect each \CN at position $i$ to each \CN at position $j$
		through a \VN if and only if $\eta_{i,j} = 1$. 

\end{enumerate}

For given $n$ and $\etab$, this construction fully specifies the
degrees of all CNs, i.e., the lengths of all component codes.  In
particular, \CNs at position $i$ have degree $d \sum_{j \neq i}
\eta_{i,j} + \eta_{i,i}(d - 1)$, where the second term arises from the
convention that we cannot connect a \gls{cn} to itself if $\eta_{i,i}
= 1$.


While our code construction is given in terms of a Tanner graph, \GPCs
have a natural array representation which we review in the following.
In general, the code array of $\Gpc$ consist of \emph{blocks}, where
$d$ is referred to as the block size.

\begin{figure}[t]
	\vspace{-0.4cm}
	\centering
	\subfloat[complete bipartite graph]{\includegraphics{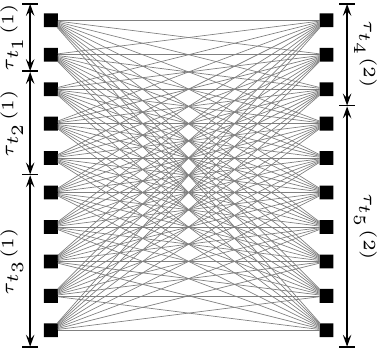}}
	\qquad
	\subfloat[code array block]{\includegraphics{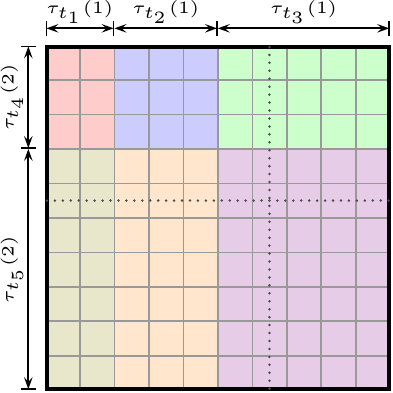}}
	\caption{Illustrations corresponding to off-diagonal entries in
	$\etab$, i.e., $\eta_{i,j} = 1$ for $i \neq j$. The distributions specifying the erasure-correcting
	capabilities are 
	$\tau_{t_1}(1) = 0.2$, $\tau_{t_2}(1) = 0.3$, $\tau_{t_3}(1) =
	0.5$, $\tau_{t_4}(2) = 0.3$, and $\tau_{t_5}(2) = 0.7$.  }
	\vspace{-0.6cm}
	\label{fig:pc}
\end{figure}

\begin{figure}[t]
	\centering
	$\,\,$
	\subfloat[complete graph]{\includegraphics{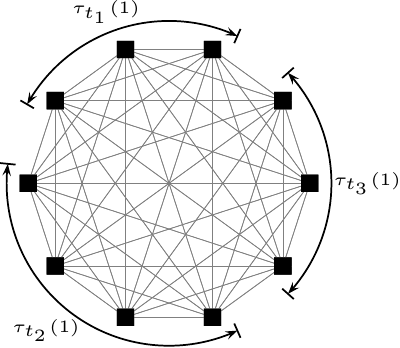}}
	\quad
	\subfloat[code array block]{\includegraphics{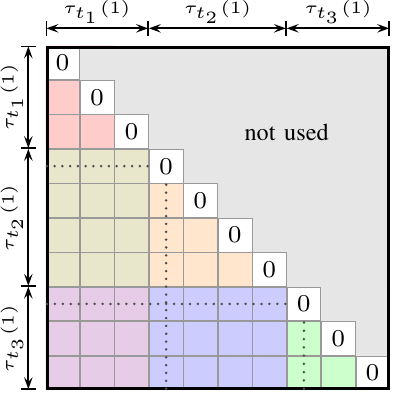}}
	\caption{Illustrations corresponding to diagonal entries in
	$\etab$, i.e., $\eta_{i,i} = 1$. The distribution specifying the erasure-correcting
	capabilities is assumed to be
$\tau_{t_1}(1) = 0.3$, $\tau_{t_2}(1) = 0.4$, and $\tau_{t_3}(1) =
0.3$.  }
	\vspace{-0.5cm}
	\label{fig:hpc}
\end{figure}

\begin{example} 
	\label{ex:pc}
	A \PC is obtained for $L = 2$ and $\etab = \left(\begin{smallmatrix}
		0 & 1\\ 1 & 0
	\end{smallmatrix}\right)$. Fig.~\ref{fig:pc}(a) shows the simplified
	Tanner graph for $n=20$ (i.e., $d = 10$), where VNs are represented
	by edges. The graph corresponds to a complete bipartite graph. The \CNs
	at the two positions correspond to ``row codes'' and ``column
	codes'', respectively. The $d \times d$ code array is shown in
	Fig.~\ref{fig:pc}(b), where colors and arrows can be ignored for
	now. One particular row/column constraint is indicated by the dotted
	lines.  \demo
\end{example}

In general, an off-diagonal entry in the lower triangular part of
$\etab$ (i.e., connecting \CNs at two different positions)
gives rise to $d^2$ \VNs and a square code array block as shown in
Fig.~\ref{fig:pc}(b). On the other hand, a diagonal entry in $\etab$
(i.e., connecting CNs at the same position) only gives rise to
$\binom{d}{2}$ \VNs. 


\begin{example} 
	\label{ex:hpc}
	Consider the case where $L = 1$ and $\etab = 1$.
	Fig.~\ref{fig:hpc}(a) shows the simplified Tanner graph
	corresponding to a complete graph for $n = 10$ (i.e., $d = 10$). The
	code is referred to as a \HPC \cite{Tanner1981, Justesen2011}. Each
	VN (i.e., each edge in the simplified Tanner graph) can be
	associated with an entry in the lower (or upper) triangular part of
	the code array block, as shown in Fig.~\ref{fig:hpc}(b).  Component
	code constraints act on L-shapes in the array, i.e., partial rows
	and columns, which include a frozen $0$-bit on the array diagonal.
	Two different code constraints are visualized in
	Fig.~\ref{fig:hpc}(b) by the dotted lines. \demo
\end{example}

The graphs and arrays in Figs.~\ref{fig:pc} and \ref{fig:hpc} are the
fundamental building blocks in the code construction. Examples of
\GPCs consisting of multiple blocks include staircase
\cite{Smith2012a} and half-braided codes \cite{Justesen2011,
Pfister2015}, which can both be seen as special cases in our
construction (see \cite{Hager2016ofc} for details). The number of
VNs---and therefore the length of $\mathcal{C}_n(\etab,
\vect{\tau})$---is
obtained by counting the bits in all code array blocks according to 
	$m = \binom{d}{2} \sum_{i=1}^L \eta_{i,i} + d^2 \sum_{1 \leq i < j
	\leq L} \eta_{i,j}.$

\subsection{Erasure-Correcting Capabilities and VN Classes}
\label{sec:vn_classes}

So far, we have only specified the lengths of the component codes
associated with the \CNs. As a last step, we assign different
erasure-correcting capabilities to the component codes. To that end,
for each $i \in [L]$, let $\vect{\tau}(i) = (\tau_1(i), \ldots,
\tau_{\tmax}(i))^\transpose$ be a probability vector of length $\tmax$
(i.e., $\vect{1}^\transpose \vect{\tau}(i) = 1$ and $\vect{\tau}(i)
\succeq \vect{0}$), where $\tau_t(i)$ denotes the fraction of
component codes at position $i$ which can correct $t$ erasures and
$\tmax$ is the maximum erasure-correcting capability.
Erasure-correcting capabilities are assigned in a consecutive fashion
as indicated by the arrows in Figs.~\ref{fig:pc} and \ref{fig:hpc}.
The collection of all probability vectors (or distributions) is
denoted by $\vect{\tau} = (\vect{\tau}(i))_{i=1}^L$. 

The class of a \CN and the corresponding component code is given
by the pair $(i,t)$, where $i$ refers to the position in the Tanner
graph and $t$ to the erasure-correcting capability. The code
construction is given in terms of \CN classes. This is somewhat
counterintuitive since typically coded bits are more tangible than
code constraints. Therefore, a description in terms of \VN classes may
be preferable. However, \VN classes are secondary in our construction:
The class of a \VN depends on the classes of the \emph{two} CNs that
it is connected to. In particular, a \VN class is defined by two
parameter pairs $(i,t)$ and $(j, t')$, i.e., four parameters in total.
The total number of VN classes is denoted by $K$.  We assume
some fixed and arbitrary indexing of the VN classes.  The number of
bits in the $k$-th VN class is denoted by $m_k$ for $k \in
[K]$.

VN classes will become important in the next section, where we discuss
how the different coded bits are transmitted over parallel channels.
Roughly speaking, the \VN class determines the protection level of the
corresponding bit. The protection level depends on the block that the
bit belongs to (i.e., the pair $(i,j)$) and the strengths of the two
associated component codes (i.e., the pair $(t, t')$). Observe that
employing component codes with different erasure-correcting
capabilities essentially subdivides each code array block into
subblocks. This is illustrated in Figs.~\ref{fig:pc} and
\ref{fig:hpc}, where arrows indicate groups of component codes with
the same erasure-correcting capability and colors represent different
VN classes. Note that we have $K = 6$ in both cases. 

\section{Channel Model and Bit Mapper} 

We assume that a codeword of $\mathcal{C}_n(\etab, \vect{\tau})$ is
transmitted over a set of $M$ parallel independent \BECs with
different erasure probabilities $p_1$, \dots, $p_\ch$. The allocation
of the coded bits to the \BECs is determined by a bit mapper. In
particular, let $\mat{A} = [a_{k,q}]$ be the bit mapper matrix of size
$K \times \ch$, where $a_{k,q}$ denotes the fraction of coded bits
from the $k$-th VN class (out of $m_k$ total bits) that are allocated
to the $q$-th \BEC. A valid bit mapper matrix is such that each row in
$\mat{A}$ is a probability vector (i.e., for each $k \in [K]$, we have
$\sum_{q =1}^{\ch} a_{k,q} = 1$ and $a_{k,q} \geq 0$ for all $q \in
[\ch]$) and, additionally, we have
\begin{align}
	\sum_{k=1}^K a_{k,q} \frac{m_k}{m} = \frac{1}{M}, \qquad \text{for all } q
	\in [\ch].
	\label{eq:cond2}
\end{align}
The condition \eqref{eq:cond2} ensures that all parallel channels are
used equally often. The set of all valid bit mapper matrices is
denoted by $\mathcal{A}$. As a baseline, we consider the ``uniform''
bit mapper $\Aunif$, where $a_{k,q} = 1/\ch$ for all $k \in [K]$ and
$q \in [\ch]$. 

The bit mapper matrix only determines the \emph{fraction} of coded
bits from a VN class that is allocated to a particular \BEC. It does
not determine which particular bit is sent through which channel. We
assume that the actual allocation is determined uniformly at random,
individually for each VN class. Such a random allocation effectively
acts as though each VN class is subject to a (potentially) different
erasure probability. In other words, one may think about transmitting
the coded bits from different VN classes through different ``virtual''
\BECs. The erasure probability for the $k$-th VN class is denoted by
$\tilde{p}_k$ and is given by 
\begin{align}
	\label{eq:mixing}
	\tilde{p}_k = \sum_{q=1}^{\ch} a_{k,q} p_q.
\end{align}
As an example, for the baseline bit mapper, all virtual \BECs have the
same erasure probability $(\sum_{q=1}^\ch p_q)/\ch$.

\section{Asymptotic Performance Analysis}

\subsection{Iterative Bounded-Distance Decoding}

We employ $\ell$ iterations of \BDD for all component codes. In
particular, each component code is assumed to correct all erasure
patterns with weight up to its erasure-correcting capability. We are
interested in characterizing the asymptotic performance of the overall
iterative decoder as $n \to \infty$. 

\subsection{Erasure Probability Scaling}
\label{sec:scaling}

For any fixed set of erasure probabilities $p_1$, \dots, $p_\ch$, it
can be shown that the decoding will fail with high probability as $n
\to \infty$. This is a simple consequence of the assumed finite
erasure-correcting capabilities of the component codes, while at the
same time the expected number of erasures per component code grows without
bounds.

In order to allow for a meaningful asymptotic analysis, we fix some
positive constants $c_1, \dots, c_\ch$ and then consider the case
where $p_1 = c_1 / n$, \dots, $p_\ch = c_\ch / n$. In other words, we
assume that the erasure probability for each \BEC decays slowly to
zero as $n \to \infty$. Due to the vanishing erasure probabilities, it
is tempting to conjecture that the decoding will now always be
successful in the asymptotic limit. However, the answer depends
crucially on the choice of $c_1, \dots, c_\ch$. Therefore, it is
instructive to think about the constants $c_1, \dots, c_\ch$ as
\emph{effective channel qualities}. 

Due to the linearity of \eqref{eq:mixing}, the bit mapper converts the
effective channel qualities for the parallel \BECs into effective channel qualities for the
virtual \BECs denoted by $\tilde{c}_1, \dots, \tilde{c}_K$, i.e., we
have $\tilde{p}_k = \tilde{c}_k / n$, where $\tilde{c}_k$ is a
weighted average of $c_1, \dots, c_\ch$. In the following, it is more
convenient to use a different indexing for the effective channel
quality $\tilde{c}_k$ associated with the $k$-th VN class. In
particular, we use the alternative notation $\tilde{c}_{t, t'}(i,j)$
instead of $\tilde{c}_k$, where $i,j,t,t'$ are the four parameters
that identify the $k$-th VN class (see Section \ref{sec:vn_classes}).  

\subsection{Density Evolution}
\label{sec:de}

For the asymptotic \DE analysis, one parameter is tracked per \CN
class as a function of the iteration number $\ell$. The parameter is
denoted by $x_{i,t}^{(\ell)}$ and its meaning is as follows. Consider
a randomly chosen erased bit that is attached to a component code with
class $(i, t)$. Then, the probability that this bit is \emph{not}
recovered after $\ell$ iterations by the component code converges
asymptotically to $x_{i,t}^{(\ell)}$. 

It can be rigorously shown that the parameter evolves as
\begin{align}
	{x}_{i,t}^{(\ell)} = \Psi_{\geq t}\left(
	\frac{1}{L} \sum_{j=1}^L \eta_{i,j} \sum_{t'= 1}^{\tmax} \tilde{c}_{t, t'}(i,j) 
	\tau_{t'}(j) {x}_{j,t'}^{(\ell-1)}
	\right),
	\label{eq:de_general}
\end{align}
where $\Psi_{\geq \tee}(x) \define 1 - \sum_{i=0}^{\tee-1}
\frac{x^i}{i!} e^{-x}$ is the tail probability of a Poisson random
variable and, initially, we have $x_{i,t}^{(0)} = 1$ for all $i$ and
$t$. Due to space constraints, we only provide some intuition behind
\eqref{eq:de_general}.\footnote{The result is a straightforward
generalization of \cite[Th.~2]{Haeger2015tit}. Compared to
the proof of \cite[Th.~2]{Haeger2015tit}, the only difference is to
take into account the different erasure probabilities for different VN
classes. However, this difference is easily captured in the
inhomogeneous random graph model \cite{Bollobas2007}.} In particular,
assume that we perform only one iteration, i.e., $\ell = 1$. Fix a
randomly chosen erased bit attached to a class-$(i,t)$ component code.
The bit will not be recovered if $t$ or more additional erasures are
attached to the same component code. To compute the corresponding
\emph{probability} of not recovering the bit, it is therefore
sufficient to characterize the distribution of the number of
additional erasures. To do so, observe that the bits of a
class-$(i,t)$ component code are split up into different sections
corresponding to different VN classes (see for example the dotted
lines in Figs.~\ref{fig:pc}(b) or \ref{fig:hpc}(b), where sections are
indicated by different colors). The number of bits per section is
given by $\tau_{t'}(j) d$ and each bit in a given section is erased
independently with probability $\tilde{c}_{t, t'}(i,j) / n$. As $n \to
\infty$, the total number of erased bits per section thus converges to
a Poisson random variable with mean $\tau_{t'}(j) d \tilde{c}_{t,
t'}(i,j) / n = \tau_{t'}(j) \tilde{c}_{t, t'}(i,j) / L$.  By
considering all sections in a class-$(i,t)$ component code (i.e.,
enumerating over all pairs $(j, t')$), we find that the total number
of erasures is Poisson distributed with mean
\begin{align}
	\frac{1}{L} \sum_{j=1}^L \eta_{i,j} \sum_{t'= 1}^{\tmax}
	\tilde{c}_{t, t'}(i,j) \tau_{t'}(j). 
\end{align}
This gives \eqref{eq:de_general} for $\ell = 1$. For subsequent
iterations, we argue as follows. Assume that at the start of iteration
$\ell$, erased bits in section $(j,t')$ have been recovered by
class-$(j,t')$ component codes independently with probability $1 -
x_{j,t'}^{(\ell-1)}$. In this case, the distribution of the
(remaining) additional erasures is still Poisson, albeit with a
reduced mean parameter according to the term inside the brackets in
\eqref{eq:de_general}.

\subsection{Decoding Thresholds}
\label{sec:thresholds}

The \DE parameters in \eqref{eq:de_general} depend on the code
parameters, the bit mapper matrix, and the values of the effective
channel qualities $c_1, \dots, c_\ch$. The latter dependence is
made explicit by writing $x_{i, t}^{(\ell)}(\vect{c})$, where
$\vect{c} \define (c_1, \dots, c_\ch)^\transpose$. We say that a
certain set of effective channel qualities is admissible if
$\lim_{\ell \to \infty} x_{i, t}^{(\ell)}(\vect{c}) = 0$ for all $i$
and $t$. This corresponds to the case where decoding will be
successful with high probability, provided that the code length and
the number of decoding iterations are sufficiently large.
Characterizing the set of all admissible effective channel qualities
then leads to a threshold region (for a given code and bit mapper). 

For some scenarios, including the coded modulation setup described in
the next section, the effective channel qualities are parametrized by
a single parameter $c > 0$. This allows us to define a one-dimensional
decoding threshold. In particular, consider the following simple
linear parametrization. Fix some constants $b_q > 0 $ for $q \in
[\ch]$ such that $(\sum_{q=1}^{\ch} b_q)/\ch = 1$. The effective
channel qualities are then given by $c_q = c b_q$. The \DE parameters
now effectively only depend on $c$, i.e., the average effective
channel quality, and we may write $x_{i, t}^{(\ell)}(c)$.  The
decoding threshold in this case is defined as
\begin{align}
	\label{eq:cthr}
	\cthr	= \sup \{ c > 0 \,|\, \lim_{\ell \to \infty} x_{i, t}^{(\ell)}(c) = 0 \text{ for all
	} i, t\}.
\end{align}

\subsection{Parallel Binary Symmetric Channels}

Typically, \GPCs are used to correct errors and not erasures.
Studying parallel \BECs is merely a trick in order to allow for a
rigorous asymptotic analysis. The problem with analyzing iterative
\BDD over \BSCs is that the component code decoders may introduce
undetected errors into the decoding process.  This happens whenever
the noise vector is such that it moves the transmitted codeword from
the correct decoding sphere into a decoding sphere corresponding to
another codeword. In that case, we say that the component code decoder
miscorrects. 

In terms of analysis, the prevailing approach in the literature (and
also the one adopted here) is to assume that a genie prevents
miscorrections. Under this assumption, the \DE analysis in Section
\ref{sec:de} also applies to the transmission over a set of parallel
\BSCs, where all previously defined erasure probabilities are now
interpreted as crossover probabilities. Furthermore, the
erasure-correcting capabilities of the component codes are interpreted
as error-correcting capabilities. 

\section{Higher-Order Modulation}

In this section, we describe how the asymptotic \DE analysis in the
previous section can be used to predict the code performance in a
coded modulation scenario, in particular a \BICM system with a
hard-decision detector. 

\begin{figure}[t]
	\centering
		\includegraphics{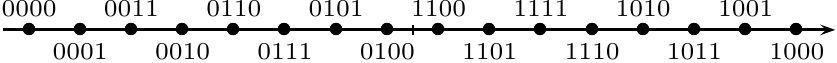}
		\vspace{-0.2cm}
	\caption{16-PAM labeled with the binary reflected Gray code.}
	\vspace{-0.4cm}
	\label{fig:16pam_brgc}
\end{figure}

Consider the \AWGN channel $Y = X + Z$, where $Z$ is a zero-mean
Gaussian random variable with variance $\sigma^2 = \E{Z^2}$. The
operating point of this channel is characterized by the \SNR $\snr =
\E{X^2}/\sigma^2$. We assume that the channel input $X$ is constrained
to a $2^\ch$-PAM constellation $\mathcal{X} = \{\pm(2i-1) \,|\, i \in
[2^{\ch-1}]\}$ which is labeled with the binary reflected Gray code.
Fig.~\ref{fig:16pam_brgc} shows the case where $\ch = 4$, i.e.,
$16$-PAM. Constellation points are selected by mapping the coded bits
(in batches of size $\ch$) to the labeling bits of the constellation.
At the receiver, a minimum-distance symbol-by-symbol detector is used
to output a hard decision on the labeling bits. 

A useful (but not necessarily exact) abstraction of this setup is to
imagine that the coded bits are transmitted over $\ch$ parallel
independent \BSCs. For sufficiently high $\rho$, the nearest-neighbor
approximation can be applied to characterize the crossover probability
of the $q$-th \BSC (corresponding to the $q$-th bit position of the
labeling) as $p_q = b_q \bar{p}(\rho)$, where $b_q = \ch
2^{q-1}/(2^{\ch} - 1)$
for $q \in [\ch]$, 
\begin{align}
	\label{eq:pbar}
	\bar{p}(\rho) =  \frac{2^M - 1}{\ch
	2^{\ch-1}}
	Q\left( \sqrt{\frac{3 \snr}{2^{2\ch}-1} } \right),
\end{align}
is the average crossover probability as a function of the \SNR $\rho$,
and $Q(\cdot)$ denotes the Q-function.   


Assuming a sufficiently large (and finite) $n$, one can use the DE
analysis in the previous section to predict the \SNR region where the
\BER performance curve of the code $\Gpc$ bends into the waterfall
behavior. In particular, we can use the linear parameterization
described in Section \ref{sec:thresholds}, where the constants $b_q$
are as defined above. The \DE analysis then gives a threshold in
terms of the average effective channel quality $c$ (see
\eqref{eq:cthr}), where we recall that the crossover probabilities are
given by $p_q = c b_q/n$. Since we also have $p_q = b_q
\bar{p}(\rho)$, we can convert any effective channel quality $c$ for a
fixed $n$ into an \SNR according to $\rho = \bar{p}^{-1}(c/n)$, where
$\bar{p}^{-1}$ is the inverse of \eqref{eq:pbar}. For example, for $M
= 4$, $c = 10.63$, and $n = 1600$, we obtain $\rho = \bar{p}^{-1}(c/n)
\approx 26.11\!$ dB. 

It is, however, important to stress that the considered \GPCs do not
have decoding thresholds in terms of the \SNR $\rho$. Rather, the
studied scaling of the crossover probabilities corresponds to the
limit $\rho \to \infty$. The calculated \SNR values for finite $n$
merely give an estimate about the \SNR region where the waterfall
region should be expected. 

\section{Bit Mapper Optimization}

As an application, we consider the problem of optimizing the bit
mapper for a fixed \GPC. For simplicity, we restrict ourselves to
irregular \HPCs where $L = 1$ and $\etab = 1$, i.e., the case where
the Tanner graph consists of only a single position (see Example
\ref{ex:hpc}). The code array corresponds to the one shown in
Fig.~\ref{fig:hpc}. To lighten the notation, position indices are
dropped. In particular, the distribution specifying different
error-correcting capabilities is denoted by $\vect{\tau} = (\tau_1,
\dots, \tau_{\tmax})$. For the optimization, we consider a
distribution $\vect{\tau}$ with two mass points according to $\tau_5 =
0.667$ and $\tau_8 = 0.333$. This gives rise to $K = 3$ different VN
classes. The indexing of the VN classes is done according to $k(5, 5)
= 1$, $k(5, 8) = 2$, $k(8,8) = 3$, where the 
notation $k(t, t')$ indicates that the $k$-th VN class corresponds to
bits that are protected by component codes with erasure-correcting
capabilities $t$ and $t'$. In the following, the signal constellation
is assumed to be $16$-PAM as shown in Fig.~\ref{fig:16pam_brgc}, i.e.,
we have $M = 4$ with $b_1 = 0.267$, $b_2 = 0.533$, $b_3 = 1.067$, and
$b_4 = 2.133$.

In order to emphasize the dependence of the decoding threshold on the
bit mapper matrix, we denote the threshold by $\cthr(\mat{A})$. For
the given parameters, we obtain for example $\cthr(\Aunif) \approx
10.13$ assuming the baseline bit mapper. Our goal is to improve the
threshold by choosing a different bit mapper. The corresponding
optimization problem can be written as $\max_{\mat{A} \in \mathcal{A}}
\cthr(\mat{A})$. One subtlety that arises for \HPCs is that the ratio
$m_k/m$ in \eqref{eq:cond2} depends on $n$, and hence also the set of
valid bit mapper matrices $\mathcal{A}$. Therefore, we replace the
``relative subblock size'' $m_k/m$ in \eqref{eq:cond2} with the
asymptotic version $\tilde{m}_k = \lim_{n \to \infty} m_k/m$, where
$\tilde{m}_k$ is either $\tau_t^2$ or $2 \tau_t \tau_{t'}$
for \HPCs.  We implemented a simple heuristic optimization solver
based on the differential evolution algorithm \cite{Storn1997} in
order to obtain a (possibly suboptimal) solution to this optimization
problem.  The found optimized bit mapper matrix is
\begin{align}
	\mat{A}^* = \begin{pmatrix}
		0.2640 &   0.1056 &   0.2984 &   0.3320 \\
    0.2976 &   0.4508 &   0.2453 &   0.0063 \\
    0.0031 &   0.0249 &   0.0746 &   0.8973
	\end{pmatrix},
\end{align}
with an improved threshold of $\cthr(\mat{A}^*) \approx 10.63$.  The
solution is not necessarily globally optimal, due to the heuristic
nature of the solver. From the optimized bit mapper matrix, one can
see for example that bits from the 3rd VN class with the highest
protection level (i.e., the last row in $\mat{A}^*$) are allocated
mainly to the labeling bit which is least reliable. 

\begin{figure}[t]
		\vspace{-0.1cm}
	\centering
		\includegraphics{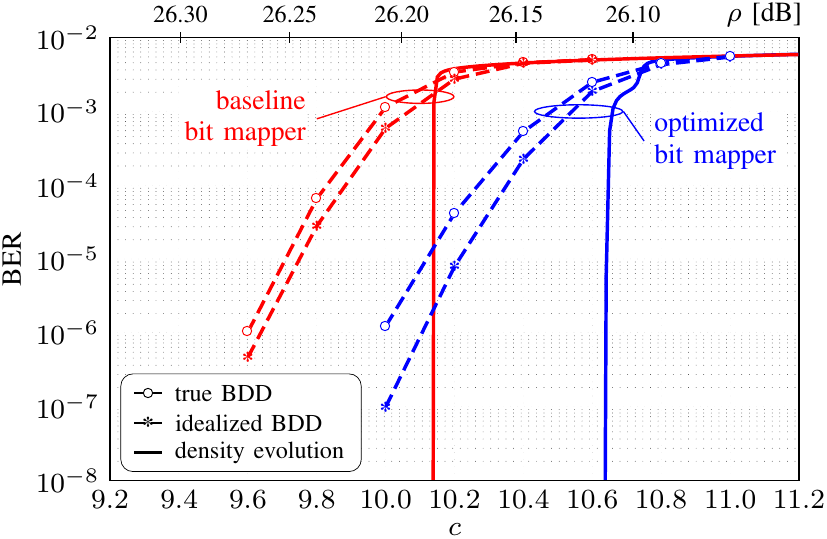}
		\vspace{-0.3cm}
	\caption{Simulation results}
		\vspace{-0.5cm}
	\label{fig:simulation2}
\end{figure}

We simulated the code $\Gpc$ for $n = 1600$ with the baseline and
optimized bit mappers\footnote{To determine the actual number of bits
to be allocated for finite values of $n$, appropriate
modifications (e.g., rounding) of the matrix $m \mat{A}^*$ are
required. } over the \AWGN channel with $\ell = 50$ iterations of
\BDD. For the chosen parameters, some rounding is required in the
sense that we use $\lfloor n \tau_5 \rfloor = 1067$ and $\lceil \tau_8
n \rceil = 533$ component codes with the two different
erasure-correcting capabilities. Note that the code has length $m =
1,279,200$ (where $m_1 = m_2 = 568,711$ ad $m_3 = 141,778$), which is
considered acceptable for fiber-optical communications due to the high
data rates.  Furthermore, assuming binary primitive \BCH codes as
component codes, it can be shown that the code rate is lower-bounded
by $R \geq 0.92$, see \cite[Sec.~VII-D]{Haeger2015tit}. 
Fig.~\ref{fig:simulation2} shows the simulation results by the dashed
lines, where dots correspond to true \BDD and stars to idealized \BDD
with no miscorrections. The \DE prediction is shown by the solid lines
and accurately predicts the waterfall regions of the two systems. By
using the optimized bit mapper, one obtains a modest improvement of
$\approx 0.06\,$dB at a \BER of $10^{-6}$. This improvement is,
however, obtained virtually ``for free'', i.e., it entails only a
reallocation of the coded bits to the signal constellation. 

\section{Conclusions and Future Work}

In this paper, we characterized the asymptotic performance of
deterministic \GPCs over parallel \BECs assuming iterative \BDD. It
was shown that the analysis accurately predicts the code performance
in a \BICM system with a hard-decision detector. We used the analysis
to optimize the bit mapper, which leads to moderate performance
improvements at almost no increased system complexity (e.g., some
additional buffering may be required). For future work, it could be
interesting to study the joint design of the code and bit mapper. 



\end{document}